\documentstyle[12pt,epsfig,epsf,psfig,rotate]{article}

\def\be{\begin{equation}}
\def\ee{\end{equation}}
\def\bea{\begin{eqnarray}}
\def\eea{\end{eqnarray}}

\begin{document}

\begin{center}
{\bf An Alternative Method to Obtain the Quark Polarization of the Nucleon}
\vskip 3mm
G. Igo
\vskip 1mm
$Department~of~Physics~and~Astronomy,~UCLA,~Los~Angeles,~California~90095$
\vskip 1mm
(October 6, 2000)
\end{center}
\begin{abstract}

An alternate method is described to extract the quark contribution to 
the spin of the nucleon directly from the
first moment of the deuteron structure function, $g^d_1$.  It is 
obtained without recourse to the use of
input on the nucleon wave function from hyperon decays involving the 
flavor symmetry parameters, F and D. The
result  for the quark polarization of the nucleon, $\Delta\Sigma_ N,$ 
is in good agreement with the values of the singlet axial
current matrix element, $a_0$, obtained from recent next-to-leading 
order  analyses of  current proton, neutron and
deuteron data. 
\newline
\newline
PAC number(s): 13.60.Hb, 13.60.-r, 11.55.Hx

\end{abstract}

In polarized deep inelastic scattering experiments, the structure 
function, $g _1(x, Q^2)$, is measured. In leading order, it is the
charge-weighted difference of the momentum distributions for quark 
helicities aligned and anti-aligned to the
longitudinally polarized nucleon,

\begin{equation}
g_1 (x,Q^2)=\frac{1}{2}\sum_i e^2_i [q^{\uparrow}_i (x,Q^2)
-q^{\downarrow}_i (x,Q^2)] = \sum_i e^2_i \Delta q_i (x,Q^2).
\label{eq:i1}
\end{equation}

The total helicity $\Delta\Sigma$  due to the quarks, on the other 
hand, depends on $\stackrel {}\int^1_0\Delta q_i ( x,
Q^2)dx\ =\Delta q_i $, the \underline {unweighted} components of the 
spin of the nucleon carried by a quark of flavor $i$ and
$\Delta\Sigma =\Delta u +\Delta d +\Delta s$.  Thus to extract 
$\Delta\Sigma$ from the measurement of the first moment of  $g_1
(x,Q^2),$ additional axial vector matrix elements, derived with the 
assumption of  the validity of isospin symmetry and SU(3)
flavor symmetry, have often been used \cite{1gluc}-\cite{4adev}. This procedure has been 
questioned since SU(3) flavor symmetry-breaking effects are
present and do not vanish for axial vector matrix elements.  The 
situation is different for the first moment of the nucleon,
$\Delta\Sigma_ N,$ defined as one-half of the sum of the neutron and 
proton helicities, in the quark parton model.  The first
moment of $ g_{1}^{N}$  is directly proportional to 
$\Delta\Sigma_ N,$  within a small term proportional to $\Delta s$.
Therefore it is possible to extract the former without a dependence 
on the SU(3) flavor symmetry parameters, F and D \cite{5clos},\cite{6caso}. In
this paper it is shown that the two methods of extracting 
$\Delta\Sigma$, one in which the flavor symmetry parameters
do not enter, and the other, the conventional approach, in which it 
is assumed that SU(3) symmetry-breaking effects can be
neglected, yield values of  $\Delta\Sigma$ in  good agreement with 
one another. It should be pointed out that in the
next-to-leading order QCD analysis \cite{7abe} by the E154 Collaboration, the 
authors do not assume SU(3) flavor symmetry to fix the
normalization of the non-singlet distributions by the axial charges 
$a_3 = F + D$  and $a_8 = 3F - D$. Their central ansatz of parameterizing the polarized parton 
distribution is that they are proportional to the
unpolarized parton distributions obtained from ref.\cite{8gluc}.
Another approach has been taken by Ohlsson and Snellman \cite{9ohl}, who employ
the broken SU(3) symmetric chiral quark model \cite{xxman} to analyze the nucleon quark spin
polarization measurements directly, without using any SU(3) symmetry assumption.
This was accomplished by fixing the symmetry-breaking parameters of the model 
with measurements by the NuSea Collaboration \cite{10haw},\cite{11pen} of the quark sea isospin
asymmetries.  They obtain a value for $\Delta\Sigma(Q^2)$ of $0.29 \pm 0.14$ at $Q^2 = 5$~GeV$^2$, in
reasonable agreement with the summary of values shown in Figure 1 below.

The first moment of the nucleon, defined as one-half of the sum of 
the first moments of the proton and neutron spin
structure functions, is expressible in terms of the first moment of 
the deuteron, $\Gamma^d_1,$ and alternately, in terms of
the axial currents, $a_0$ and $a_8$ \cite{9shur}:

\begin{equation}
\Gamma^N_1 = \Gamma^d_1 /(1- \frac{3}{2}\omega_D) =\int_{0}^{1} 
g^d_1 (x,Q^2)dx / (1 -
\frac{3}{2}\omega_D) = (\frac{1}{36} a_8 C_{ns} + \frac{1}{9} a_0 C_s).
\label{eq:i2}
\end{equation}
Here  $\omega_D$ is the D-state probability of the deuteron and the 
quantities $C_{ns}$  and $C_s$  are
$Q^2$-dependent coefficient functions. In the case of the proton 
(neutron) there is additional dependence on the axial
current, $a_3$ ; i.e., for the proton:

\begin{equation}
\Gamma^p_1 (Q^2) = \int^1_0 g^p_1 (x,Q^2)dx = (\frac{1}{12} a_3 + 
\frac{1}{36} a_8)C_{ns} +
\frac{1}{9} a_0 C_s.
\label{eq:i3}
\end{equation}

If polarized gluons do not contribute to the nucleon spin, 
the singlet and
non-singlet matrix elements of the axial current are expressible in 
terms of $\Delta u,\Delta d$  and $\Delta s$ .  Assuming
isospin invariance and SU(3) flavor symmetry, they can also be 
expressed in terms of the weak hyperon decay constants, F and D, as\\

\begin{equation}
a_0 = \Delta u + \Delta d + \Delta s = \Delta \Sigma ~~~,
\end{equation}
\begin{equation}
a_3 = \Delta u - \Delta d = F + D 
\end{equation}
and
\begin{equation}
a_8 = \Delta u + \Delta d - 2 \Delta s = 3F - D .
\end{equation}
\\

The non-singlet coefficient function $C_{ns}$  and the singlet 
coefficient function $C_s$  appearing in Equations \ref{eq:i2} and
\ref{eq:i3} have been calculated  in the modified minimal subtraction 
$\overline {MS}$ scheme \cite{9lari},\cite{10lar} to  third order for three quark
flavors. The non-singlet coefficient  is found  to be,

\begin{equation}
C_{ns} = 1 - \frac{\alpha_s(Q^2)}{\pi} - 3.58 (\frac{\alpha_s(Q^2)}{\pi})^2
-20.22(\frac{\alpha_s(Q^2)}{\pi})^3.
\label{eq:i4}
\end{equation}
where $\alpha_s(Q^2)$ is the strong coupling constant. The  singlet 
coefficient function $C_s$ has been evaluated in
two forms in the $\overline {MS}$ scheme, one of which yields a 
$Q^2$-dependent $a_0$ in equations \ref{eq:i2} and \ref{eq:i3} and one, used
below, which provides the high $Q^2$  limit to $a_0$ (The conclusions 
reached below are not affected substantially by
this choice), is

\begin{equation}
C_s = 1-0.3333\frac{\alpha_s (Q^2)}{\pi} - 0.5495 
(\frac{\alpha_s(Q^2)}{\pi})^2 \ \ .
\label{eq:i5}
\end{equation}

In the $\overline {MS}$  scheme, gluons do not contribute to the 
first moment of $g_1$ and $a_0 = \Delta\Sigma,$ and the former is
independent of the factorization scheme. In the Adler-Bardeen (AB) 
scheme \cite{11bal}, the invariant quantity, $a_0$ is not equal to
$\Delta\Sigma,$
and because of the axial anomaly \cite{12adl} also involves the gluon spin 
structure  function, $\Delta G(Q^2)$,

\begin{equation}
\ a_0 =\Delta \Sigma\ - \frac{3}{2 \pi} \alpha_s (Q^2)\Delta G (Q^2).
\label{eq:i6}
\end{equation}

The values of $a_0,$ extracted from the analysis by the SMC 
Collaboration \cite{4adev} of the
world data in the two schemes, are consistent with one another within 
their experimental errors. In the extraction
of $a_0,$ from the world data for proton, deuteron and neutrons, the 
normalization of the non-singlet quark densities
are fixed using the neutron and hyperon  $\beta$-decay constants and 
assuming SU(3) flavor symmetry in the SMC analysis.
The relation between the matrix element  $a_3,$ and the neutron 
$\beta$-decay constants, $g_A \slash g _V = F + D,$ relies
only on the assumption of isospin invariance. However in order to 
relate $a_8$ to the semi-leptonic hyperon decay constants,
F and D,  SU(3)  flavor symmetry is assumed and hence  conclusions on 
$a_0$  depend on its validity.  SU(3)  symmetry-breaking
effects do not vanish at first order for axial vector matrix elements 
\cite{13gai}. It has been suggested that in order to reproduce
the experimental values of F and D, the quark-parton model requires 
large relativistic corrections which depend on
the quark masses. Since the s-quark mass is much larger than that of 
u and d quarks, these corrections should
break  SU(3)  symmetry. Similarly, the relations between the baryon 
magnetic moments predicted by  SU(3) are badly
broken \cite{14dzi}. The uncertainty in $a_8$  propagates into  $a_0$.  For 
instance, the value obtained for $a_8$  from a
leading-order 1$/N_c$  expansion is much smaller that the value based 
on the SU(3) analysis. The use of a smaller value of $a_8$
causes $a_0$  to become larger \cite{xxdai},\cite{9ohl}.  Symmetry breaking effects have been 
calculated under various assumptions leading to variations
in the quantity  3F - D  differing by a factor of two \cite{15jaf}-\cite{19son}. Lipkin 
\cite{20lip}  criticizes the use of SU(3) symmetry  in
extracting the singlet axial current from the data, and maintains 
that the use of F and D can only be considered
fudge factors with no physical basis that  should work for the axial 
current.

In view of the reservations discussed in the last paragraph about the 
use of the SU(3) flavor symmetry parameters in the extraction of $a_0,$ 
it is of interest to compare 
with the results from the alternate  method
described here.  As noted above, in the quark-parton model, 
the first moment $\Gamma^{N}_{1}$  of the nucleon
structure function,
$g{^N_1} = \frac{\ g{^p_1+g{^n_1}}}{\ 2},$ can be expressed, up to a 
small correction factor depending on $\Delta s,$ wholly in
a term proportional to $\Delta\Sigma,$ i.e. expressible without 
invoking SU(3) symmetry and as a consequence, F and D.
We write

\newpage

\begin{eqnarray}
\Gamma_1^N &=& \Gamma^d_1/(1 - \frac{3}{2}\omega_D)=\frac{1}{36} 
[(\Delta u + \Delta d - 2 \Delta s)C_{ns}
+ 4 (\Delta u + \Delta d + \Delta s)C_s]\nonumber\\
&=& \frac{1}{36} [(C_{ns} + 4C_s)\Delta \Sigma] - \frac{1}{12} C_{cs} \Delta s.
\label{eq:i7}
\end{eqnarray}

\begin{figure}[hbt]
\epsfig{file=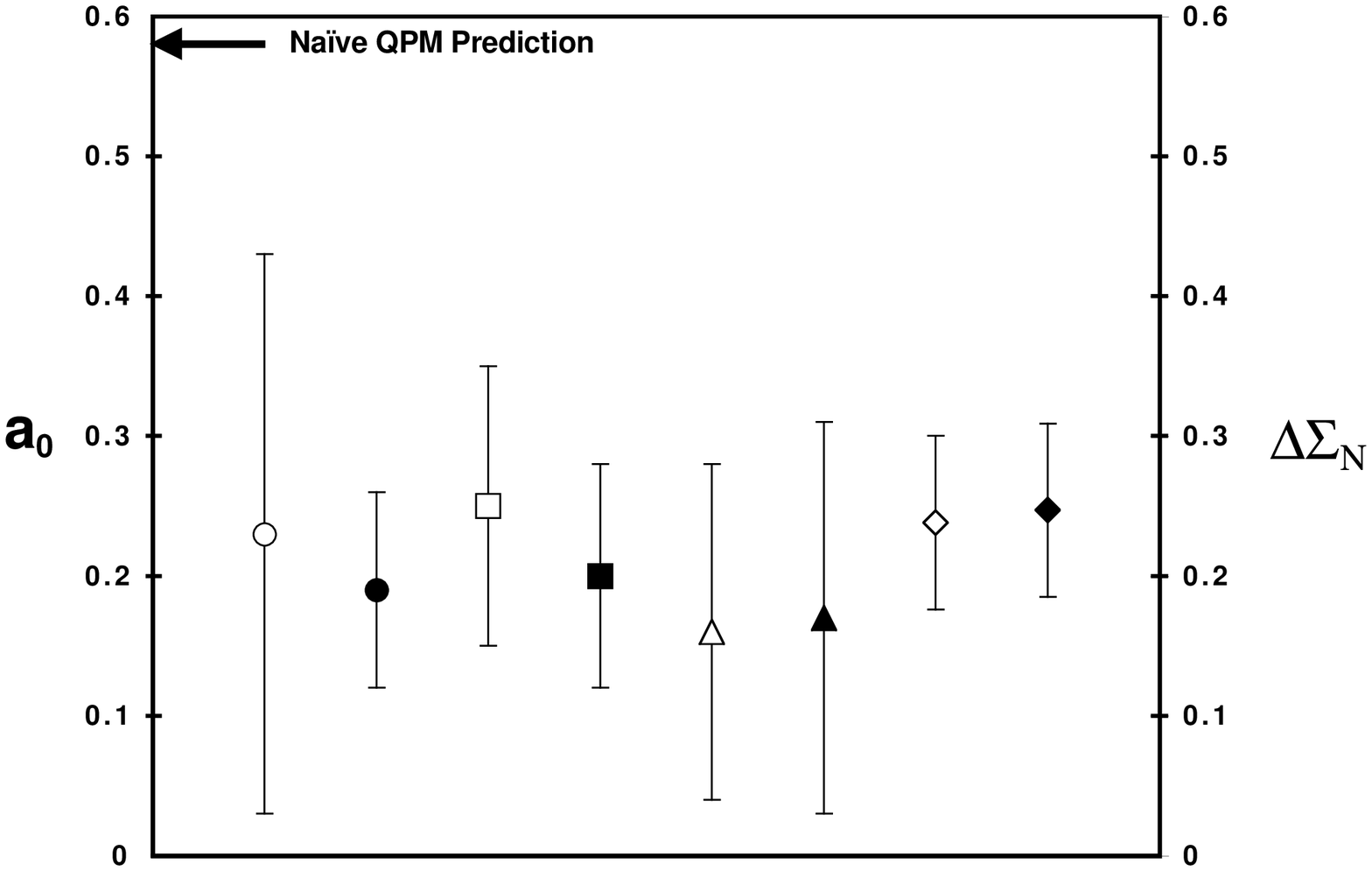, width=\textwidth}
\caption{The singlet axial current matrix element, $a_0$, obtained from the 
analysis of world data by the SMC Collaboration[4] using the $\overline {MS}$ scheme 
(open circle), using the AB scheme (solid circle); same by the E154 Collaboration[7]
using the $\overline {MS}$ scheme (open square), using the AB scheme (solid square).
The quantity $\Delta\Sigma_{N}$ obtained from the first moments of the deuteron structure
function measurement by the SMC Collaboration[23] (open triangle) and
world data[4](solid triangle); same with measurements by the E155 Collaboration[22]
using the low $x$ extrapolation of the E154 Collaboration[7] (open
diamond) and using the low $x$ extrapolation of the SMC Collaboration[4] (solid
diamond).}
\label{ig001}
\end{figure}

Neglecting the last term in Equation \ref{eq:i7} for the moment,

\begin{equation}
\Delta \Sigma_N = 36 \Gamma_1^d /[(C_{ns} + 4C_s)(1 - \frac{3}{2}\omega_D)] ,\\
\end{equation}
thus expressing $\Delta\Sigma_ N$ without explicit dependence on the 
F and D parameters. In Figure \ref{ig001}, $\Delta\Sigma_ N$
at $Q^2$ = 5 GeV$^2$  is plotted using the values for $\Gamma_{1}^{d}$ obtained by the 
E155 Collaboration at 5 GeV$^2$ \cite{22ant} and at 10 GeV$^2$
by the SMC Collaboration \cite{23ade}. These are in excellent agreement with 
the analysis of the world data for $\ a_0$ at $Q^2$ = 5
GeV$^2$ by the E154 Collaboration \cite{7abe} using the $\overline {MS}$ 
scheme and the AB scheme as illustrated in Figure~\ref{ig001}. In the
former scheme  $\ a_0$ is equal to $\Delta\Sigma$; in the AB scheme, 
by Equation \ref{eq:i6}. Figure~\ref{ig001} also gives the values of $a_0$ 
obtained by the SMC Collaboration in their analysis \cite{4adev} of the 
world data for a starting value of 1 GeV$^2$  in both 
the $\overline {MS}$ and AB  schemes. These are also in good agreement with the 
values of $\Delta\Sigma_N$ obtained in the 
analysis described in this paper. Inclusion of the term
dependent on $\Delta s$  in equation \ref{eq:i7} reduces the values of 
$\Delta\Sigma_ N,$ plotted in Figure~\ref{ig001} by approximately one-half
of their experimental errors when  $\Delta s$  is set equal to -0.08, 
consistent with the analysis by the E143 Collaboration
\cite{25abe} of their data. They obtained values for $\Delta s$ of $-0.08 \pm 
0.03, -0.10 \pm 0.03, -0.07\pm 0.04.$

In summary, it is shown in this note that the singlet axial current 
matrix element, $a_0,$ obtained by the E154 \cite{7abe}
and the SMC \cite{4adev} Collaborations in their respective analyses of the 
world data on proton, neutron and deuteron
structure functions with the latter using SU(3) flavor  asymmetry and 
the  parameters F and D are quite
consistent with $\Delta\Sigma_N,$  obtained directly from the 
deuteron spin structure functions without the assumption of SU(3)
flavor symmetry. This result provides further substantiation that the 
observed discrepancy shown in Figure \ref{ig001} of the
naive QPM prediction for $a_0$ \cite{21ell} and experiment is not an 
artifact due to a significant breakdown of SU(3) flavor
symmetry.

\newpage

\begin{center}
{\bf Acknowledgements}
\end{center}
This work was supported by the U.S. Department of Energy.
I would like to thank S. Trentalange for a careful reading of the manuscript.

\end{document}